\documentclass[11pt,twocolappendix,iop]{emulateapj}
\newcommand{\myemail}{herve.bourdin@roma2.infn.it}

\usepackage{amsmath}
\usepackage{natbib,txfonts} 
\usepackage{graphicx} 

\def\K{\mathrm{K}}

\def\ho{\mathrm{H}_{0}}

\def\lx{\mathrm{L}_{\mathrm{X}}}
\def\kt{\mathrm{kT}} 
\def\kto{\mathrm{kT_o}} 
\def\abo{\mathrm{Z_o}} 
\def\ab{\mathrm{Z}} 
\def\T{\mathrm{T}}
\def\kev{\mathrm{keV}}

\def\w{\mathrm{W}}

\def\nh{\mathrm{N}_\mathrm{H}}
\def\nho{\mathrm{N}_\mathrm{H,o}}

\def\sx{\mathrm{\Sigma}_\mathrm{x}}

\def\stwo{\mathrm{S}_2}
\def\sone{\mathrm{S}_1}
\def\cfone{\mathrm{CF}_1}
\def\cftwo{\mathrm{CF}_2}

\def\a521{A521}
\def\xmm{\textit{XMM-Newton}~}
\def\chandra{\textit{Chandra}~}
\def\rosat{\textit{ROSAT}~}

\def\f{\mathrm{F_{\mathrm{evt}}}}
\def\s{\mathrm{F_{\mathrm{ICM}}}} 

\def\b{\mathrm{F_{\mathrm{bck}}}}

\def\nevt{\mathrm{n}_\mathrm{evt}} 
\def\nf{\mathrm{N}_\mathrm{evt}} 
\def\ns{\mathrm{N}_\mathrm{ICM}}
 
\def\nb{\mathrm{N}_\mathrm{bck}}

\def\ea{\mathrm{E}}
\def\vf{\mathrm{e}}
\def\Theta{\mathrm{Theta}}

\def\fig{Figure~}
\def\tab{Table~}
\def\equ{Equation~}
\def\equs{Equations~}
\def\part{Section~}
\def\parts{Sections~}

\shorttitle{Shock heating of A521}
\shortauthors{Bourdin et al.}

\begin{document}
\title{Shock heating of the merging galaxy cluster A521}

\author{H. Bourdin\altaffilmark{1}, P. Mazzotta \altaffilmark{1,2}, M. Markevitch\altaffilmark{3,4}, S. Giacintucci\altaffilmark{4,5}, and G. Brunetti\altaffilmark{6}}
%\email{herve.bourdin@roma2.infn.it}

\altaffiltext{1}{Dipartimento di Fisica, Universit\`a degli Studi di Roma `Tor
  Vergata', via della Ricerca Scientifica, 1, I-00133 Roma, Italy; \myemail}
\altaffiltext{2}{Harvard Smithsonian Centre for Astrophysics, 60 Garden Street, 
Cambridge MA, 02138, USA}
\altaffiltext{3}{NASA Goddard Space Flight Center, Code 662, Greenbelt, 
MD 20771, USA}
\altaffiltext{4}{Joint Space-Science Institute, University of Maryland, College Park, MD, 20742-2421, USA}
\altaffiltext{5}{Astronomy Department, University of Maryland, College Park, 
MD 20742, USA}
\altaffiltext{6}{INAF - Istituto di Radioastronomia, via Gobetti 101, I-40129 Bologna, Italy}

\begin{abstract}

\a521 is an interacting galaxy cluster located at z=0.247, hosting 
a low frequency radio halo connected to an eastern radio relic. 
Previous \chandra observations hinted at the presence of an X-ray 
brightness edge at the position of the relic, which may be a shock front. 
We analyze a deep observation of \a521 recently performed with \xmm in 
order to probe the cluster structure up to the outermost regions covered 
by the radio emission. The cluster atmosphere exhibits various brightness 
and temperature anisotropies. In particular, two cluster cores appear 
to be separated by two cold fronts. We find two shock fronts, one that 
was suggested by \chandra and that is propagating to the east, and another 
to the southwestern cluster outskirt. The two main interacting clusters 
appear to be separated by a shock heated region, which exhibits a spatial 
correlation with the radio halo. The outer edge of the radio relic 
coincides spatially with a shock front, suggesting this shock is responsible 
for the generation of cosmic ray electrons in the relic. The propagation 
direction and Mach number of the shock front derived from the gas 
density jump, $\mathrm{M} = 2.4\pm0.2$, are consistent with 
expectations from the radio spectral index, under the assumption of Fermi I 
acceleration mechanism.

\end{abstract}


\section{Introduction}

Collisions between massive galaxy clusters are the most energetic
events in the present universe. Part of the kinetic energy released
during these collisions is dissipated through supersonic shock fronts 
propagating in the intracluster medium (ICM) and turbulent motions. 
While heating the thermal component of the ICM, shocks and turbulence may 
also accelerate (or reaccelerate) relativistic particles \citep[e.g.,][]{Cassano_05, 
Hoeft_07, Ryu_03, Pfrommer_06, Brunetti_07, Vazza_09}.
Radio observations probe these complex mechanisms through the detection of
diffuse synchrotron emission from the ICM, in the form of giant radio halos,
Mpc-scale radio emission in the cluster central regions, 
and radio relics, sharp-edged radio sources in the cluster periphery
\citep[e.g.,][for recent reviews]{Ferrari_08, Cassano_09, Venturi_11, Brunetti_11}

Observable as sharp X-ray brightness and temperature discontinuities, 
few shock fronts have been detected so far, because they can only be visible 
in the brightest cluster regions and in particularly favorable projections 
\citep[][]{Markevitch_02, Markevitch_05, Russell_10, Finoguenov_10, Macario_11}. Peripheral 
radio relics are believed to be shock fronts that propagated far outside 
the X-ray bright region, while still accelerating (or re-accelerating) electrons, 
which produce radio emission and quickly cool after the shock passes, resulting in a
characteristic narrow feature \citep[e.g.,][]{Ensslin_98,Van_Weeren_11}.
The physics of giant radio halos is probably more complex. 
Radio halos plausibly result from
the (re)-acceleration and transport 
of relativistic particles in large turbulent regions of the ICM, 
although many aspects of the mechanisms generating 
radio-emitting electrons remain unclear \citep[e.g.,][for a recent review]{Brunetti_11}.
Sharp radio edges (and radio relics) are frequently observed at the border 
of giant radio halos, suggesting a possible link between merger shocks 
and the generation of turbulence in the ICM \citep[e.g.,][]{Markevitch_10,Macario_11}.

\a521 is a moderately distant (z=0.247) 
and X-ray luminous 
\citep[$\lx = (5.2+1.2) \times 10^{37}\, \w$,][]{Arnaud_00}
\footnote{$10^{37}\, \w \equiv 10^{44} \mathrm{erg}.\mathrm{s}^{-1}$; 
X-ray luminosity in the 0.1-2.4 keV band has been corrected for luminosity 
distance assuming $\ho = 70$ km s$^{-1}$ Mpc$^{-1}$, and $\Lambda = 0.7$.} 
galaxy cluster, presenting several signatures of dynamical activity. 
As revealed already in \rosat images, its X-ray and optical components 
appear spatially segregated, with an N--S bimodality of the X-ray 
emission \citep{Arnaud_00}, and a more complex galaxy number density 
distribution revealing two NW/SE and NE/SW major 
elongations \citep{Ferrari_03}. 
As further shown from Chandra data analysis, the ICM in \a521 exhibits 
an irregular thermal structure with indications for gas heating at 
the interface between the two main gas components \citep{Ferrari_06}. 
A521 exhibits a giant radio halo that is the prototype of the class of ultra-steep 
spectrum radio halos \citep{Brunetti_08}.  The halo being spatially connected to a radio relic, 
\a521 provides us with an ideal test case to investigate the effects 
of shocks on the properties of thermal and nom-thermal components 
of the ICM and their connection with giant radio halos.
In this merging cluster, a shock has been suggested by the presence 
of an X-ray brightness edge on the SE side of the cluster, coinciding 
with the edge of the radio relic \citep{Giacintucci_08}. 
The larger-scale cluster radio halo shows a very steep synchrotron
spectrum supporting a picture where relativistic electrons are
stochastically re--accelerated by the non-linear interaction with 
turbulence in the ICM \citep{Brunetti_08}.

The present article will focus on the analysis of a deep 
observation of \a521 recently performed with \xmm, with particular 
goal to probe the ICM structure up to the outermost regions 
covered by the cluster radio halo, and radio relic. 
After discussing data preparation and analyzes issues 
in \parts\ref{sect:data_preparation} and \ref{sect:data_analysis}, 
we present the various X-ray brightness and temperature features 
revealed by this observation in \part\ref{sect:icm_thermodynamics}. 
We comment on the interplay between thermal and non-thermal components 
of the ICM in \part\ref{sect:non_thermal_icm}. 
Unless otherwise noted, any energy distribution is normalized as a 
probability density function, while confidence ranges on individual 
parameter estimates are 68 $\%$. In the following, intra-cluster distances are 
computed as angular diameter distances, assuming a $\Lambda$-CDM cosmology 
with $\mathrm{H}_\mathrm{0} =70~\mathrm{km}~\mathrm{s}^{-1}~\mathrm{Mpc}^{-1}$, $\Omega_{\mathrm{M}} = 0.3$, $\Omega_{\Lambda} = 0.7$. 
Given these assumptions, an angular separation of 1 arcmin 
corresponds to a projected intra-cluster distance of 232.5 kpc.

\begin{table*}[ht]
\caption{Effective exposure time of each \xmm{}-EPIC observation. \label{pointing_params_tab}}
\begin{center}
\begin{tabular}{ccccc}
\tableline\tableline
XMM-Newton  & Centre coordinates & MOS1  effective  &  MOS2  effective &  PN  effective \\
obs. IDs & & exposure time (ks) & exposure time (ks) & exposure time (ks)  \\
\tableline
0603890101 (S) &  04h54m22.00s -10$^\circ$16'30." & 15.7 (50.8  \%)  & 15.7  (68.9 \%) & 12.8  (36.1 \%) \\
0603890101 (U) &  04h54m22.00s -10$^\circ$16'30." & 64.4 (77.3  \%)  & 64.5  (74.6 \%) & 57.6  (52.5 \%) \\
\tableline \\
\multicolumn{5}{l}{\textbf{Note.} The fraction of the useful exposure time after solar-flare filtering is shown in brackets.} \\
\end{tabular}
\end{center}
\end{table*}

\begin{figure}[t]
  %\resizebox{\hsize}{!}{\includegraphics{./figures/a521_fov.eps}}
  \resizebox{\hsize}{!}{\includegraphics{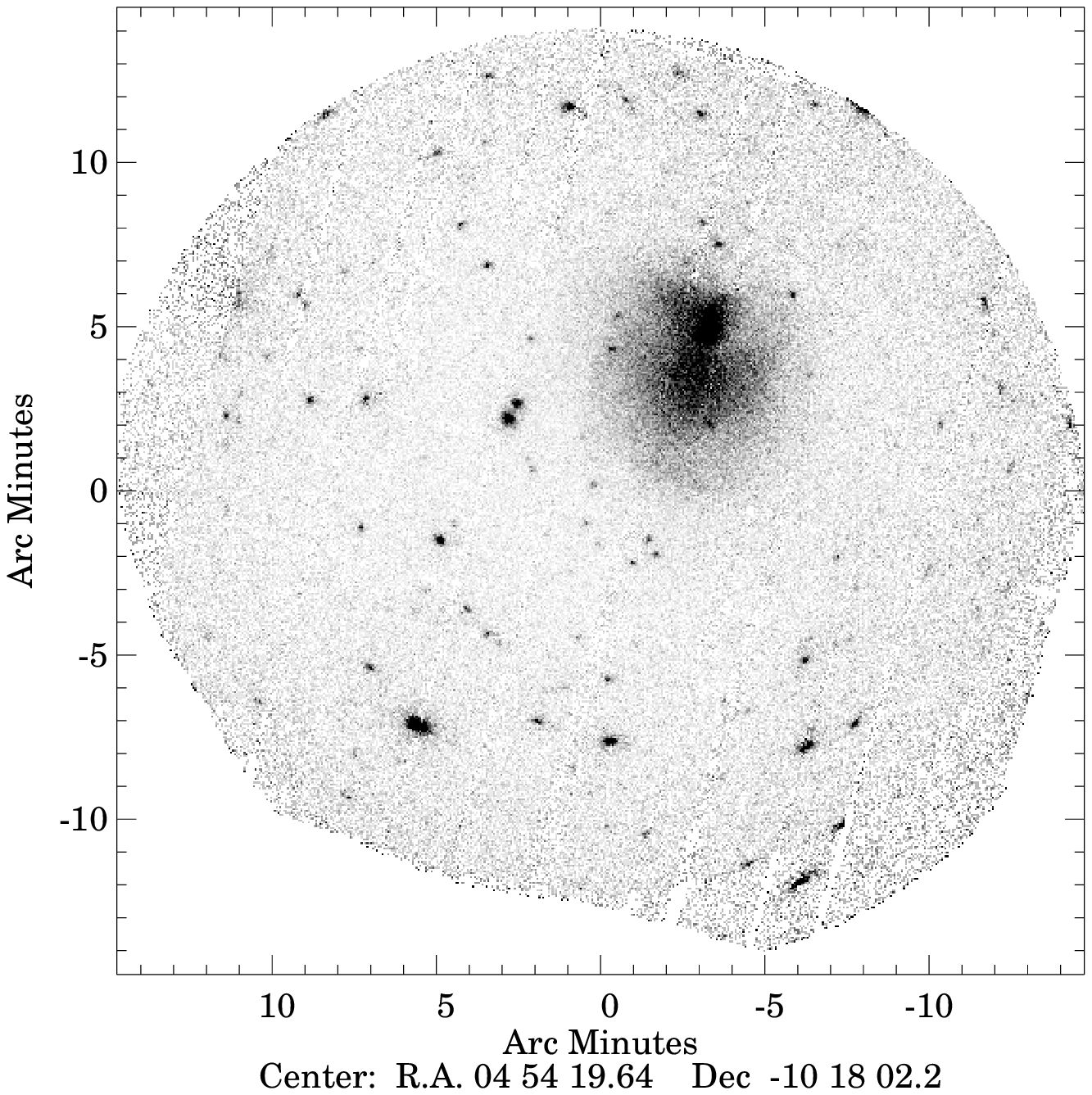}}
  \caption{EPIC \xmm exposure of A521.\label{a521_fov}}
\end{figure}

\begin{figure}[t]
  %\resizebox{\hsize}{!}{\includegraphics{./figures/a521_zoom_cxb_model.eps}}
  \resizebox{\hsize}{!}{\includegraphics{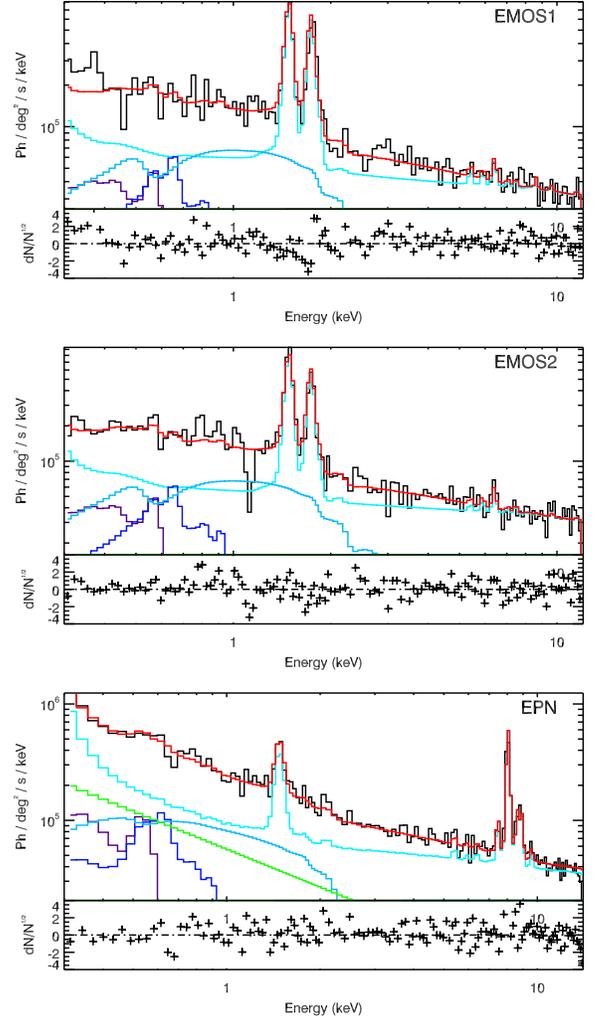}}
  \caption{Background spectrum observable in the A521 outskirts. 
  Light blue: particle background. Cyan blue: Cosmic X-ray Background emission.
Blue and Violet: TAE emission (kT1 = 0.099 keV, kT2 = 0.248 keV,
see \citet{Kuntz_00}, and details in Sect. \ref{sect:background}). Green: Residual soft proton emission. 
  Red and black: overall fit and data set.\label{cxb_model_fig}}
\end{figure}

\section{Observations and data preparation}{\label{sect:data_preparation}}

The EPIC-\xmm{} data set is a dual observation of \a521, performed with focal aim point and central EPIC-MOS CCDs located in the SE cluster outskirts (see \fig\ref{a521_fov}). In order to remove the contribution of soft proton flares, we filtered the histogram of the photon arrival times through a temporal wavelet analysis. A summary of the `good' exposure time remaining on each of the three EPIC cameras is provided in \tab\ref{pointing_params_tab}. The average good exposure time is about 75 ks. 

In order to perform imaging and spatially resolved spectroscopy, we binned photons in sky coordinates $(k,l)$ and energy $(e)$, matching the angular and spectral resolution of each focal instrument.  To map the surface brightness of extended sources, these photon counts may have to be normalized for spatial and spectral variations of the telescope effective area and detector exposure times. We thus associate an `effective exposure' array, $\ea(k,l,e)$, 
to the photon event cube. Expressible e.g. in $\mathrm{s}.\mathrm{deg}^2$, $\ea(k,l,e)$  is computed as a linear combination of CCD exposure times, $t_{\mathrm{CCD}}(k,l,p)$, related to individual observations $p$, with local corrections for useful CCD areas, $ a_{\mathrm{CCD}}(k,l,p)$, Reflexion Grating Spectrometer (RGS) transmissions\footnote{EPIC-MOS detectors share a common optical path with the RGS}, $tr_{\mathrm{RGS}}$, and mirror vignetting factors $a_{\mathrm{CCD}}(k,l,p)$\footnote{Information about these instrumental effects have been obtained from the \xmm-EPIC Current Calibration Files (CCFs)}.

\begin{eqnarray}
\ea(k,l,e) = \sum_{p=1}^\K t_{\mathrm{CCD}}(k,l,p) && \times ~\Delta a_{\mathrm{mirror}}(k,l,e,p) \nonumber \\
&& \times ~ tr_{\mathrm{RGS}}(k,l,e,p) \nonumber \\
&& \times ~ a_{\mathrm{CCD}}(k,l,p)  
\label{eff_exposure_equ}
\end{eqnarray}

\section{Data analysis}{\label{sect:data_analysis}}

\subsection{Background noise modeling\label{sect:background}}

The cluster emissivity must be separated from an additive, spatially extended and mostly stationary background noise including false photon detections due to charged particle-induced and out-of-time events, but also the Cosmic X-ray Background (CXB), and some Galactic foreground components.

The XMM EPIC background is dominated by
the particle component, which is modeled from observations performed in the 
Filter Wheel Closed (FWC) position during revolutions 230 to 2027 as for the 
EPIC-MOS cameras, and 355 to 1905 as for the EPIC-PN camera. Following an approach proposed in
e.g. \citet{Kuntz_08} or \citet{Leccardi_08} this model
sums a quiescent continuum to a set of florescence emission lines
convolved with the energy response of each detector. 
It is completed with a residual emission associated with 
soft protons, presently only detectable in the case 
of the EPIC-PN camera and modelled as a power-low 
spectrum normalized to 1.4 cts $\mathrm{deg}^{-2}~\mathrm{s}^{-1}$ in the 
0.5--1. $\kev$ band. To account for two different spectral shapes in the soft and hard bands, 
the quiescent continuum is modelled as the product of a power law with an 
inverted error function increasing in the soft band. We set the emission line 
energies to the values reported in \citet{Leccardi_08}, while the soft proton residual is 
modelled using an additional power law. Presumably due to differences in the collecting areas 
of the imaging and readout detector regions, the EPIC-MOS quiescent continuum exhibit a small
emissivity gradient along the RAWY CCD coordinate, which has been measured
and taken into account in the model. Because the fluorescence lines exhibit a more complex 
spatial variation \citep{Lumb_02, Kuntz_08}, we modelled the 
emissivity distribution of the most
prominent lines \footnote{Namely the Al, Si and Cu, Ni complexes as
for the EPIC-MOS and EPIC-PN cameras, respectively.} from the wavelet
filtering of a set of FWC event images in narrow energy intervals around each 
line.

Secondary background components include the Cosmic X-ray background 
and Galactic foregrounds. Being associated with real photon detections, these components
are corrected for the effective exposure. The Cosmic X-ray background is modelled with an 
absorbed power law of index $\gamma = 1.42 $ \citep[see, e.g.,][]{Lumb_02}, while the Galactic
foregrounds are modelled by the sum of two absorbed thermal components
accounting for the Galactic Ôtransabsorption emissionÕ \citep[TAE; kT1 = 0.099 keV and kT2 = 0.248 keV, see][]{Kuntz_00}. 
We estimate emissivities of each of these components from a ``joint-fit" of all background components in a region of the 
field of view located beyond the boundary of X-ray emission in the SE cluster outskirt, but covered by the central MOS CCDs 
(see also \fig\ref{cxb_model_fig}). %At a 1 keV energy, 
This estimates yields  13.4, 28.0 and 29.5 cts.$\mathrm{m}^{-2}\mathrm{deg}^{-2}.\mathrm{s}^{-1}$ in the 0.5--1. $\kev$ band as for the two transabsorption and CXB components, respectively ($\chi^2$/d.o.f. = 1.23). Our background model, $\nb(k,l) \sum_{e}\b(k,l,e)$, eventually includes a contribution for the EPIC-PN out-of-time count rate, which is estimated in each energy band as 6.3 \% of all photon counts registered along the CCD columns.

\subsection{Spectroscopic and surface brightness measurements}

To estimate average ICM temperatures, $\kt$, and  metal 
abundances, $\ab$, along the line of sight and for a given location of the 
field of view $(k,l)$, we add a source emission 
spectrum to the background model, and fit the spectral shape of the resulting 
function, $\nf(k,l)~\f(\kt,\ab,\nh,e)$, to the photon 
energy distribution registered in the energy band (0.3--12 keV):

\begin{eqnarray}
  && \nf(k,l)~\f(\kt,\ab,\nh,e)  \nonumber \\
  &=& \ea(k,l,e) \times \ns(k,l)~\s(\kt,\ab,\nh,e) \nonumber \\
  && + \nb~\b(k,l,e).
%\label{global_spectra_equ}
\label{temperature_estimate}
\end{eqnarray}

In this modeling, the source emission spectrum $\s(\kt,\ab,\nh,e)$ assumes a redshifted and $\nh$ absorbed emission modelled from the Astrophysical Plasma Emission Code \citep[APEC,][]{Smith_01}, with the element abundances of \citet{Grevesse_98} and neutral hydrogen absorption cross sections of \citet{Balucinska-Church_92}. The $\nh$ value has been fixed to $4.9 \times 10^{24} m^{-2}$, from measurements obtained near A521 in the Leiden/Argentine/Bonn Survey of Galactic HI \citep{Kalberla_05}. It is altered by the mirror effective areas, filter transmissions and detector quantum efficiency$^8$, and convolved by a local energy response matrix M(k,l,e,e') computed from response matrixes files (RMF) tabulated in detector coordinates in the \xmm-EPIC calibration data base. \footnote{EPIC response matrixes are computed from canned RMFs corresponding to the observation period provided by the \xmm Science Operation Centre.}

To compute images and radial profiles of the intra-cluster gas distribution, we estimate a cluster surface brightness map, $\sx(k,l)$, from photon counts registered in a given energy band, and corrected for effective exposure and additive background. Assuming an average ICM energy distribution, $<\s(e)>$, we define $\sx(k,l)$ as a function of an effective exposure map, $\vf(k,l)=\sum_{e}<\s(e)>\ea(k,l,e)$:

\begin{equation}
		\sx(k,l) = \frac{\sum_{e}{\nevt(k,l,e)-\nb(k,l) \sum_{e}\b(k,l,e)}}{\vf(k,l) }.
	\label{brightness_estimate}
\end{equation}

All parameters of $<\s(e)>~=\s(\kto,\abo,\nho,e)$ are practically determined from spectral fitting of the main cluster emission spectrum: $\kto = 6.7~\kev$, $\abo=.4\ab_{\odot}$, $\nho=4.9 \times 10^{24} m^{-2}$, while $\sx(k,l)$ is estimated in a ``soft" energy band ([.5--2.5] $\kev'$) in order to lower the dependence of $<\s(e)>$ on $\kto$.

\subsection{Surface brightness, ICM density and temperature profiles\label{radial_profiles_sect}}

In the following, surface brightness and temperature profiles have been extracted within cluster sectors 
oriented approximately along the surface brightness gradients. We derived the radial surface brightness $\sx(r)$ by averaging the surface brightness $\sx(k,l)$ of \equ(\ref{brightness_estimate}) in each profile annulus composed of N pixels $(k,l)$, as follows:

\begin{equation}
	\sx(r) = \frac{1}{N}\sum_{k,l} \sx(k,l),
\end{equation}

The background contribution $\nb(k,l) \sum_{e}\b(k,l,e)$ being estimated within a much larger area of the field of view 
than any sector annulus used to derive $\sx(r)$, we neglected any systematic uncertainty related to its modeling and estimated the variance on $\sx(r)$ from a weighted mean of the local Poisson fluctuations in $\sx(k,l)$:

\begin{equation}
\sigma_{\sx}(r)^2  = \frac{1}{N} {\sum_{k,l}\frac{\sigma_{\nevt}(k,l)^2}{\vf(k,l)^2}} = \frac{1}{N} {\sum_{k,l}\frac{{\nevt}(k,l)}{\vf(k,l)^2}}
\end{equation}

Projected temperatures $\kt(r)$ and associated confidence interval $\delta \kt(r)$ have been computed 
within each annulus by fitting a uniform emission model to the data set. To do so, we averaged the emission 
models of \equ(\ref{temperature_estimate}) associated with each pixel $(k,l)$ of the annulus, and estimated 
the model parameters $\kt(r),\ab(r),\nh(r)$, via a $\chi^2$ minimization.

These brightness and temperature profiles have been used to model the underlying density and temperature of the ICM, assuming spherical symmetry of the cluster atmosphere in the vicinity of the features of interest. This was undertaken by projecting and fitting parametric distributions of the three-dimensional (3D) emission measure $^{12}$, $n_p n_e$, and temperature, $\T(r)$, to the observed profiles. In this modeling, projected brightness profiles are convolved with the \xmm focus Point Spread Function (PSF), while projected temperatures are computed assuming the `spectroscopic-like' weighting scheme proposed in \citet{Mazzotta_04}. In \part\ref{shock_fronts}, the ICM emission measure \footnote{More precisely, the ICM emission measure per volume unit.} and temperature profiles across two shock fronts have been modelled by step-like distributions with a common jump radius $r_j$:

\begin{equation}
  [n_p n_e](r) = \left\{ \begin{matrix}
    \mathrm{D}_{n}^2 n_o^2~(r/r_{j})^{-2\eta_1}, ~r<r_{j} \\
    n_o^2~(r/r_{j})^{-2\eta_2}, ~r>r_{j}
  \end{matrix} \right. ,
  \label{npne_cf_equ}
\end{equation}

\begin{equation}
  \T(r) = \left\{ \begin{matrix}
    \mathrm{D}_{\T} \T_o, ~~r<r_{j} \\
	\T_o, r>r_{j} \end{matrix} \right. .
  \label{t3D_cf_equ}
\end{equation}

\begin{figure*}[ht]
  %\resizebox{\hsize}{!}{\includegraphics{./figures/a521_xmaps_pink.eps}}
  \resizebox{\hsize}{!}{\includegraphics{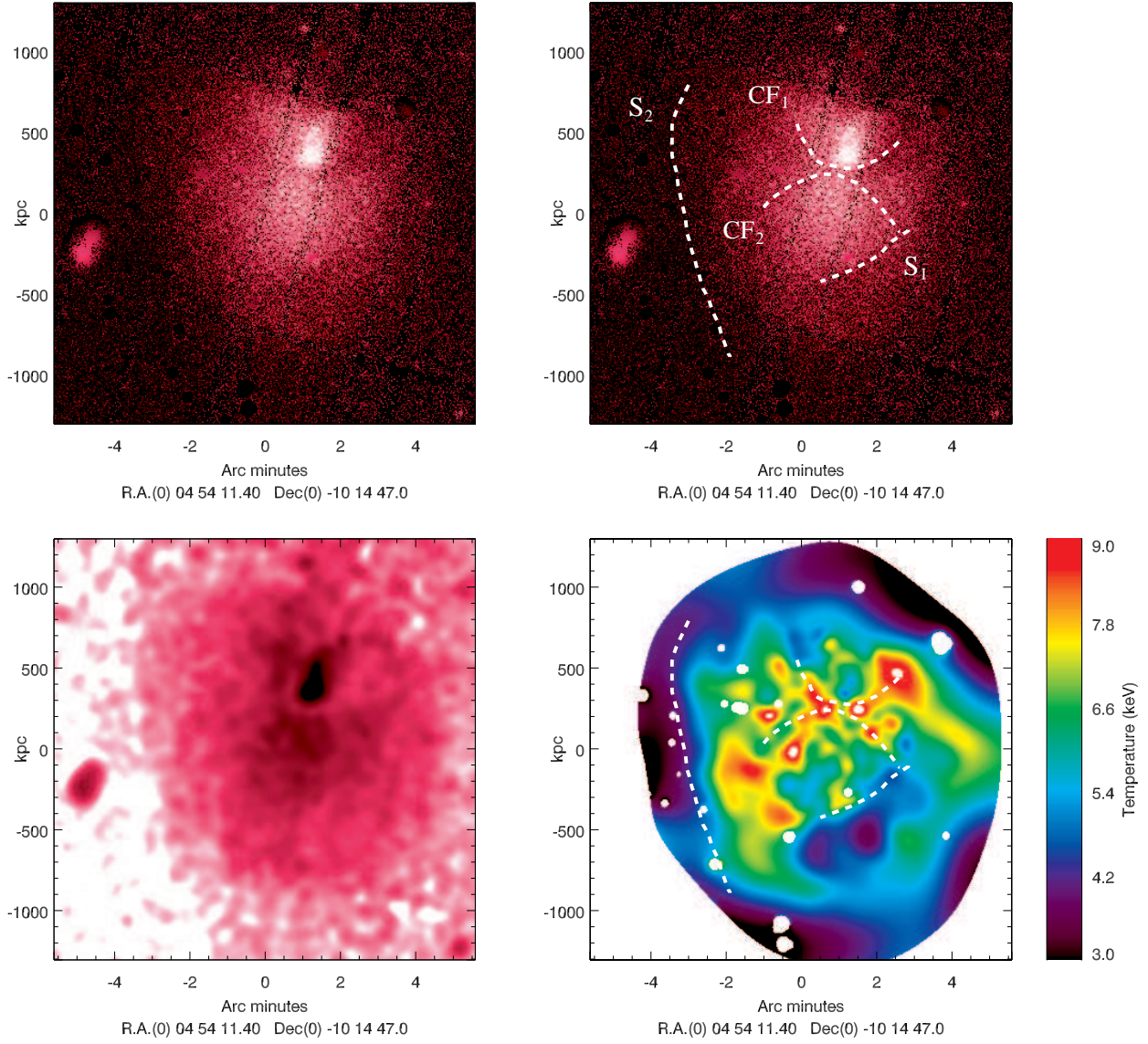}}
  \caption{EPIC \xmm observation of \a521. { Top panels:} Photon rate image in the .5-2.5 keV band. Photon counts in 
  this image have been corrected for spatially variable effective area, background flux and wavelet detected point-like 
  sources. { Bottom-left panel:} Anisotropic details in the ICM emissivity map. These details have been enhanced from subtraction of a wavelet denoised map to the photon rate (further details are provided in Sect.~\ref{imaging_sect}) { Bottom-right panel:}  ICM temperature map obtained from wavelet spectral-imaging. Prominent brightness jumps are indicated by dashed lines on the photon rate image and temperature map.
  \label{xmaps_fig}}
\end{figure*}

\subsection{Imaging and spectral-imaging\label{lxkt_maps}}

\subsubsection{Imaging\label{imaging_sect}}

An image of the cluster is presented on the top panels of \fig\ref{xmaps_fig}. To obtain this image, $\sx(k,l)$, 
we corrected the EPIC-\xmm raw photon image for spatially variable effective area and background flux, following \equ(\ref{brightness_estimate}). The point-like sources have also been modelled by means of an isotropic undecimated B3-spline wavelet analysis \citep[see e.g.][]{Starck_07}, and subtracted from the image.

A map of anisotropic details in the ICM structure is shown on the bottom-left panel of \fig\ref{xmaps_fig}. To create this image, we subtracted a wavelet filtered map of the photon rate, $\sx(k,l)$, from the photon rate itself, then smoothed
the residual image with a gaussian function of typical width $\mathrm{FWHM}=20~\mathrm{arcsec}$. The wavelet 
filtering has been performed by means of a soft 3$\sigma$ thresholding of B3-spline wavelet coefficients, the significance thresholds being directly computed from the raw ---Poisson distributed--- photon map, following
the multiscale variance stabilization scheme introduced in \citet{Zhang_08}.

\begin{figure}[]
\begin{center}
%\resizebox{\hsize}{!}{\includegraphics{./figures/a521_cf_jumps.eps}}
\resizebox{\hsize}{!}{\includegraphics{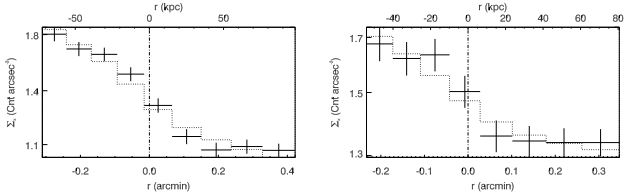}}
\caption{Projected gas brightness measured across two cluster sectors intercepting the brightness jumps $\cfone$ and $\cftwo$. 
The projection of a step-like gas density distribution (\equ\ref{npne_cf_equ}) convolved with the \xmm PSF is superimposed as a dotted line, assuming density jump amplitudes of $1.7 \pm .1$ and $2.1 \pm .1$ as for $\cfone$ and $\cftwo$, respectively.
\label{a521_cf_fig}}
\end{center}
\end{figure}

\begin{figure*}[t]
%\resizebox{.5\hsize}{!}{\includegraphics{./figures/a521_middle_arc_shock_4t_rjump.eps}}
%\resizebox{.5\hsize}{!}{\includegraphics{./figures/a521_outskirt_shock_rjump_psf.eps}}
\resizebox{.5\hsize}{!}{\includegraphics{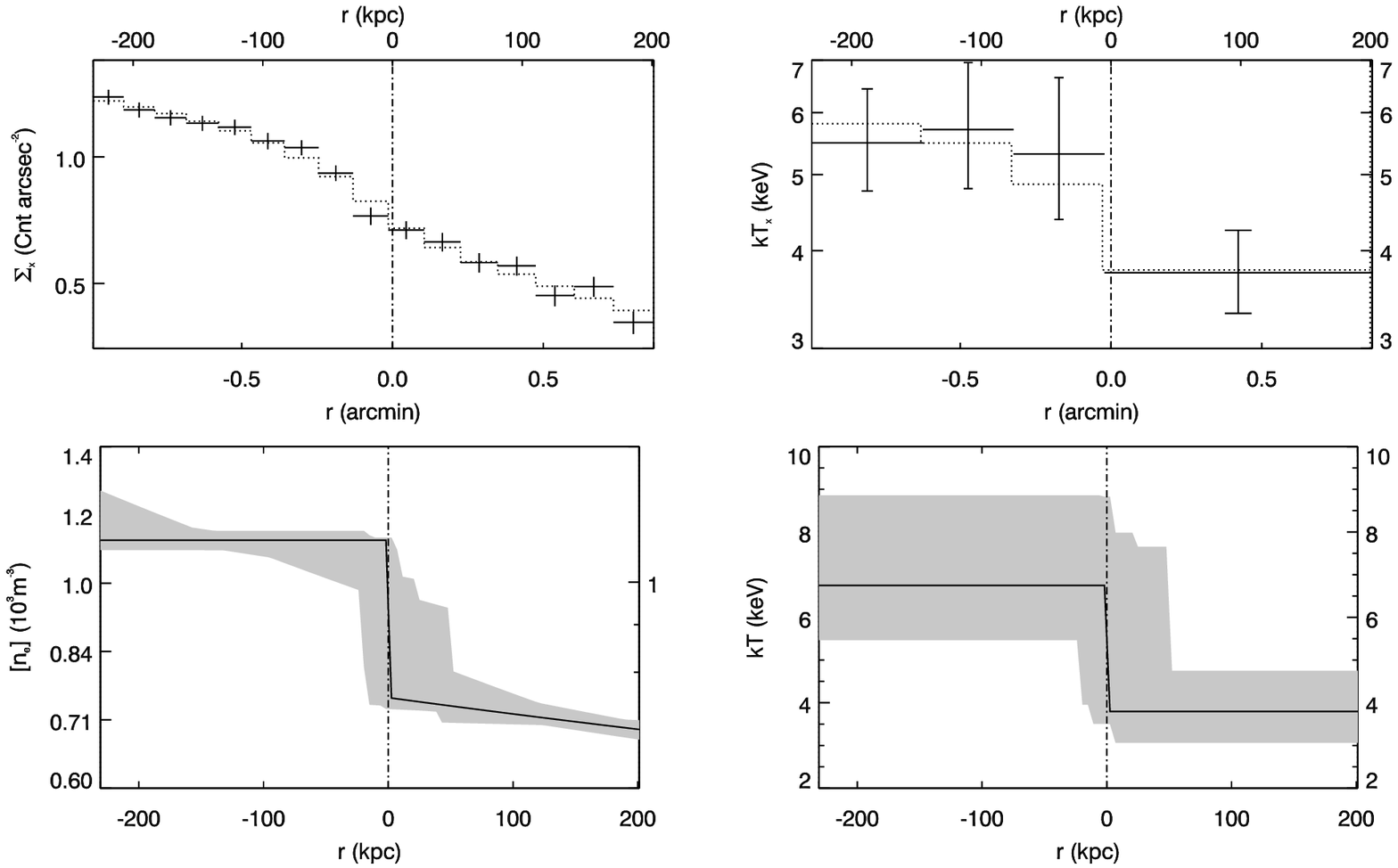}}
\resizebox{.5\hsize}{!}{\includegraphics{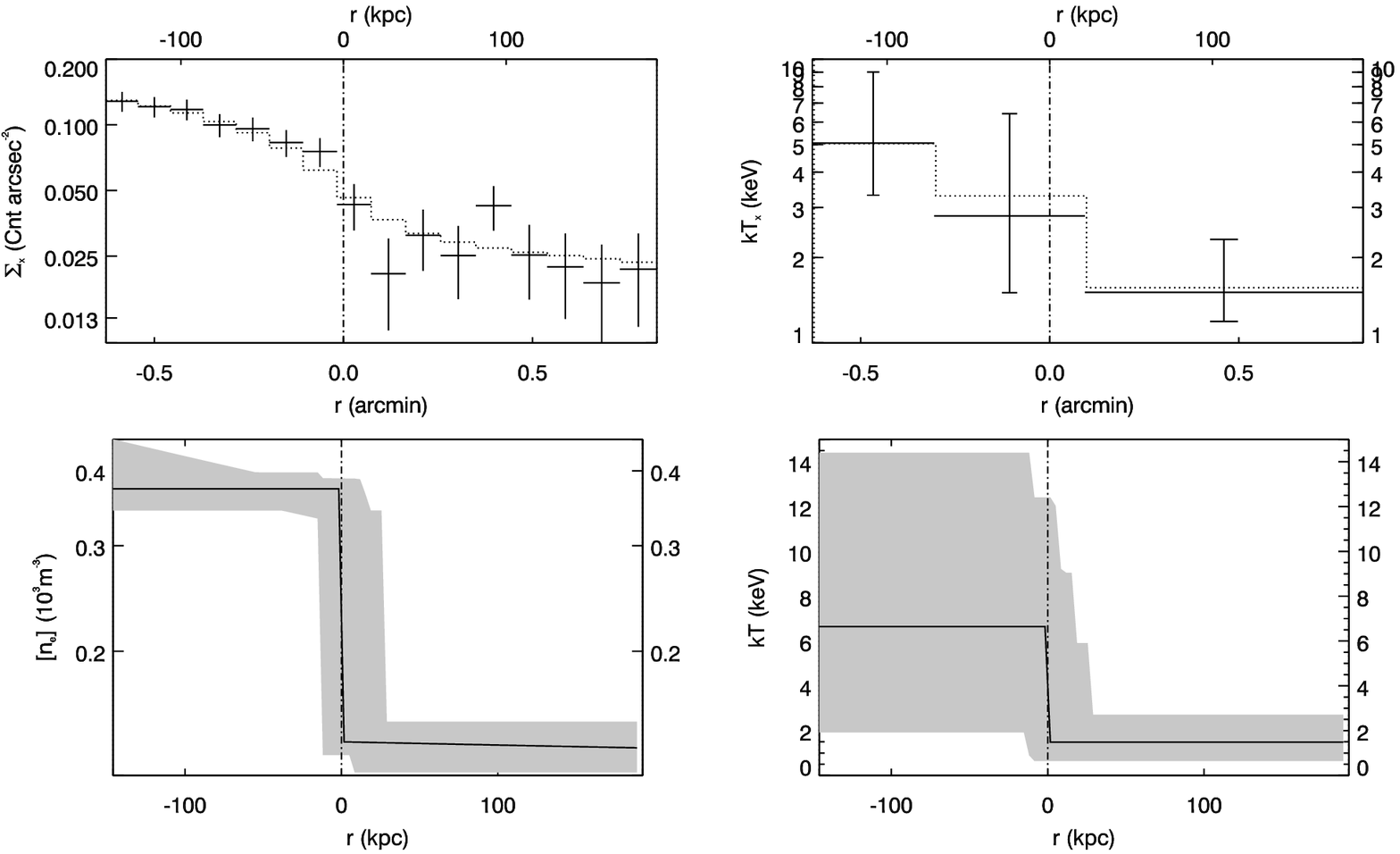}}
  \caption{{ Top panels:} Projected gas brightness and temperature profiles measured across two cluster sectors intercepting the brightness jumps $\sone$ and $\stwo$, as shown on \fig\ref{shock_region_fig}. { Bottom panels:} ICM density 
and temperature profiles modelled as step-like 3D distributions matching the projected profiles (see also \equs~\ref{npne_cf_equ} and \ref{t3D_cf_equ}). Dispersions on these profiles have been estimated from random realizations 
of the data set and corresponding models, each profile envelope delimiting 68~\% of the realizations with closest $\chi^2$ distance from the original data set. The projection of these distributions is reported as a dotted line on the projected profiles.\label{a521_shocks_fig}}
\end{figure*}

\subsubsection{Spectral-imaging}

In order to map the ICM temperature in A521, we used the
EPIC-XMM-Newton data set and applied the spectral-imaging
algorithm detailed in \citet{Bourdin_04} and \citet[][, hereafter B08]{Bourdin_08}. Following this algorithm, a set of temperature arrays $\kt(k,l,a)$ with associated fluctuations  $\sigma_{kt}(k,l,a)$ are first computed on various analysis scales $a$, then convolved by complementary high-pass and low-pass analysis filters in order to derive wavelet coefficients. The wavelet coefficients are subsequently thresholded according to a given confidence level in order to restore a de-noised temperature map. Here, the signal analysis have been performed over 6 dyadic scales within an angular resolution range of $\delta a$ = [1.7 -- 110] arcsec. This was undertaken by averaging the emission modelled by \equ(\ref{temperature_estimate}) within overlapping meta-pixels $(k,l,a)$, and computing the $\kt(k,l,a)$ and $\sigma_{kt}(k,l,a)$ arrays by means of a likelihood maximization. The resulting ICM temperature map shown in \fig\ref{xmaps_fig} was then obtained from a B2-spline wavelet analysis (see B08 for details) with coefficients thresholded to the 1$\sigma$ confidence level.
    
\begin{figure}[ht]
  %\resizebox{\hsize}{!}{\includegraphics{./figures/a521_cross_diagram2.eps}}
  \resizebox{\hsize}{!}{\includegraphics{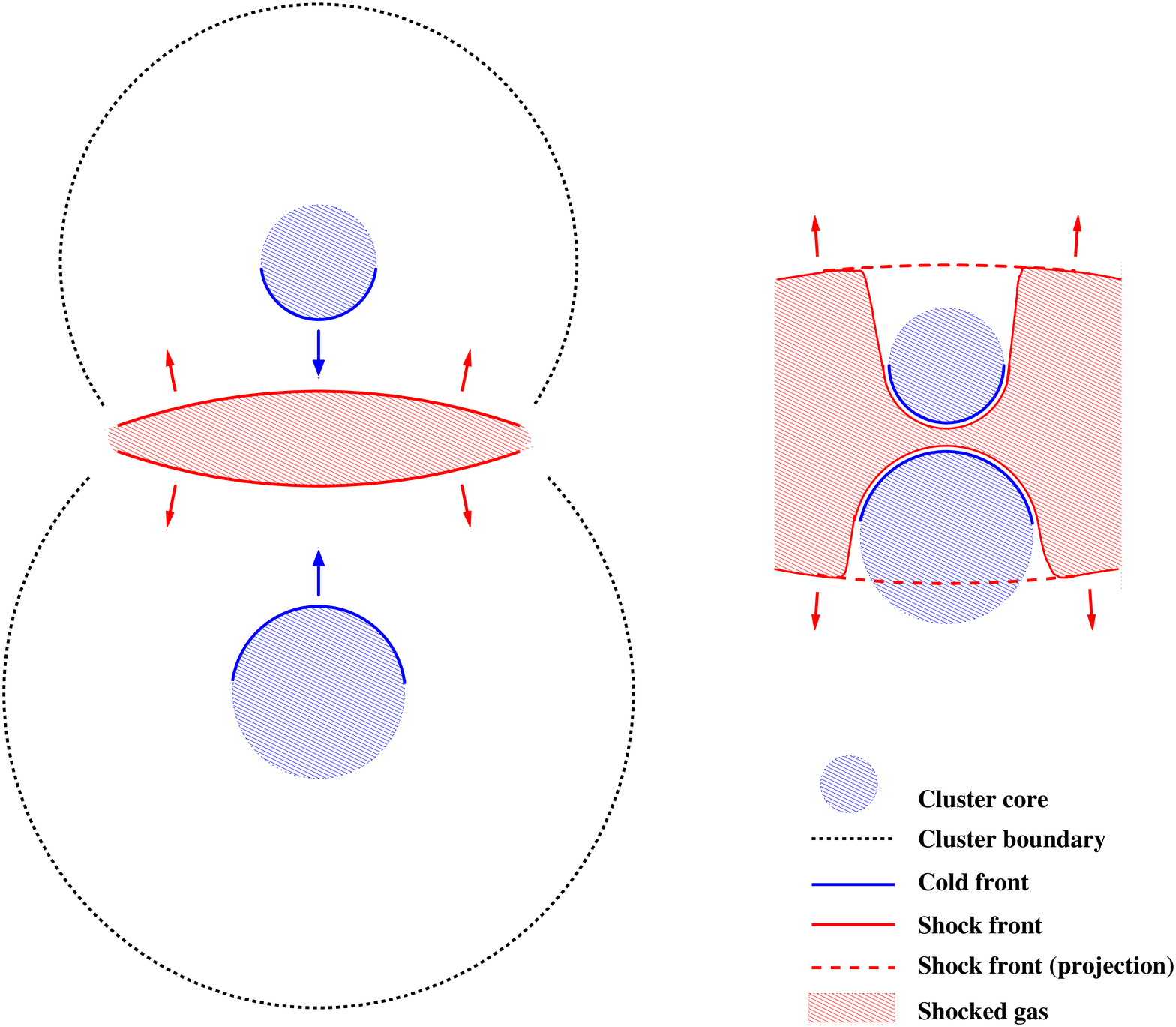}}
  \caption{Tentative interpretation of the ICM thermal and entropy structure observed in the central region of \a521. { Left:} Early stage of a two-cluster merger: the cluster boundaries start to collide and develop two shock fronts propagating within the densest regions of each cluster. In the meanwhile, the cluster develop two cold fronts while pushing the higher entropy gas away from their interacting region. { Right:} The shock fronts have now propagated to the most external regions of the interacting clusters, but could not penetrate the two cool cores. A shocked gas region with high entropy remains at the interface between the two cold fronts.\label{cross_diagram_fig}}
\vspace{.25cm}
\end{figure}

\begin{figure}[t]
  %\resizebox{\hsize}{!}{\includegraphics{./figures/a521_labelled_shock_regions.eps}}
  \resizebox{\hsize}{!}{\includegraphics{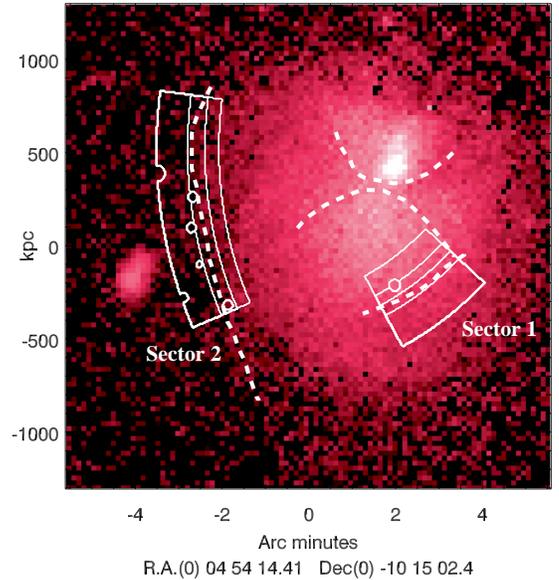}}
  \caption{Photon rate image of \a521 extracted in the .5-2.5 keV band. The image has been re-binned to a 6.8 arcsec
  angular resolution in order to enhance the brightness jumps $\sone$ and $\stwo$. The two annular sectors show the two cluster regions where temperature and brightness profiles of \fig\ref{a521_shocks_fig} have been extracted.\label{shock_region_fig}}
\end{figure}

\begin{table*}[]
\caption{Estimated cluster photon counts within the .3--5. keV band, in the regions shown on \fig\ref{shock_region_fig}
 \label{shock_region_tab}}

\begin{center}
\begin{tabular}{lrrrrrrr}
\tableline\tableline
&\multicolumn{4}{c}{Sector 1} & \multicolumn{3}{c}{Sector 2} \\
Detector & Region 1 & Region 2 & Region 3 & Region 4 & Region 1 & Region 2 & Region 3  \\
\tableline
EPIC-MOS1 & $985 ~( 79.4 \%)$ & $518 ~( 86.1 \%)$ & $592 ~( 88.9 \%)$ & $ 879 ~( 90.4 \%)$ & $140 ~( 19.6 \%)$ & $ 173 ~( 37.0 \%)$ & $232 ~( 49.8 \%)$ \\
EPIC-MOS2 & $1016~( 80.4 \%)$ & $ 552~( 87.6 \%)$ & $ 578~( 89.4 \%)$ & $868~( 91.0 \%)$ & $87~( 14.9 \%)$ & $144~( 36.4 \%)$ & $232~( 49.8 \%)$ \\
EPIC-PN      & $2086~( 74.0 \%)$ & $1083~( 82.3 \%)$ & $ 1146~( 86.4 \%)$ & $ 1557~( 87.3 \%)$ & $ 192~( 13.6 \%)$ & $284~( 29.2 \%)$ & $420~( 42.9 \%)$  \\
\tableline\\
\multicolumn{5}{l}{\textbf{Note.} The fraction of the total counts is shown in brackets.} \\
\end{tabular}
\end{center}
\end{table*}

\section{ICM thermodynamics}{\label{sect:icm_thermodynamics}}

\subsection{Intra-cluster gas brightness and thermal structure}

The X-ray photon image of \fig\ref{xmaps_fig} reveals us the complex morphology of the intra-cluster gas in \a521. On large scales, a northern subcluster with comet shape is apparently falling on the main component. The photon image also reveals the strongly irregular morphology of the surface brightness, presenting various edges indicated with dashed lines. Some of these brightness jumps have been enhanced in the bottom-left image of anisotropic details. They are also noticeable on the surface brightness profiles of \fig~\ref{a521_cf_fig} and \ref{a521_shocks_fig}. At the interface between the two main interacting cluster components, we observe two bow-shaped brightness jumps, $\cfone$ and $\cftwo$,  joining each other to form a low brightness cross-shaped feature. A third brightness jump with higher curvature radius, $\sone$, is crossing the southern cluster component from SE to NW, while a fourth one, $\stwo$,  is visible at the South-east cluster outskirts.

The ICM temperature map of \fig\ref{xmaps_fig} is strongly irregular, and presents various noticeable features. The northern sub-cluster is clearly cool ($\kt \simeq 4.5 \kev$). The interacting region separating this cool core from the main cluster to the South appears hotter  ($\kt > 7 \kev$) and strongly disturbed. The cross-shaped brightness depression observable on the photon image seems to coincide with a hot cross ($\kt \simeq 9 \kev$), in particular, along the brightness jump $\cfone$. The southern part of the main cluster is cooler ($\kt \simeq 4 \kev$) than the interacting region, in particular, to the South of the brightness jump $\sone$. 
 
Bringing together the brightness and temperature maps of \fig\ref{xmaps_fig}, we observe that the two brightness jumps $\cfone$ and $\cftwo$ are associated with temperature increases as the brightness decreases, while the brightness jumps $\sone$ and $\stwo$ are associated with a temperature decrement. The two jumps $\cfone$ and $\cftwo$ are thus likely to be cold fronts separating the densest parts of the two sub-clusters from their interacting region, while $\sone$ and $\stwo$ are probably shock fronts propagating outwards from the colliding clusters.

\subsection{Cold fronts and shock heating at the interface between two interacting sub-clusters\label{shock_heating}}

One of the most striking features seen in our data is a cross-shaped brightness depletion separating the two colliding sub-clusters. This feature also corresponds to a temperature and entropy enhancement, in particular just outside the two cold fronts, $\cfone$ and $\cftwo$. What we see is probably shocked gas with high entropy being squeezed by
the converging cool core remnants and flowing around the densest part of the two interacting clusters, without penetrating the two cold fronts. The projected layer of shocked gas would thus exhibit maximal temperature and entropy near the two cold fronts, where it is tangentially intercepted by the line of sight. This shocked gas layer might also partly overlay in projection the main sub-cluster from its boundary delineated by cold front $\cftwo$, to the southern brightness jump, $\sone$. A possible interpretation for the origin of this hot gas flow is illustrated on \fig\ref{cross_diagram_fig}. Originally located at the cluster boundary (if there is one), the high entropy gas may have been shock heated between the two clusters starting to interact. It would now expand over the cluster atmosphere, following shock fronts presently propagating outside the cluster cores. One of these shock fronts might be observed to the South of the main cluster as the brightness and temperature jump, $\sone$. 

\begin{figure*}[ht]
%\resizebox{\hsize}{!}{\includegraphics{./figures/a521_x_radio.eps}}
\resizebox{\hsize}{!}{\includegraphics{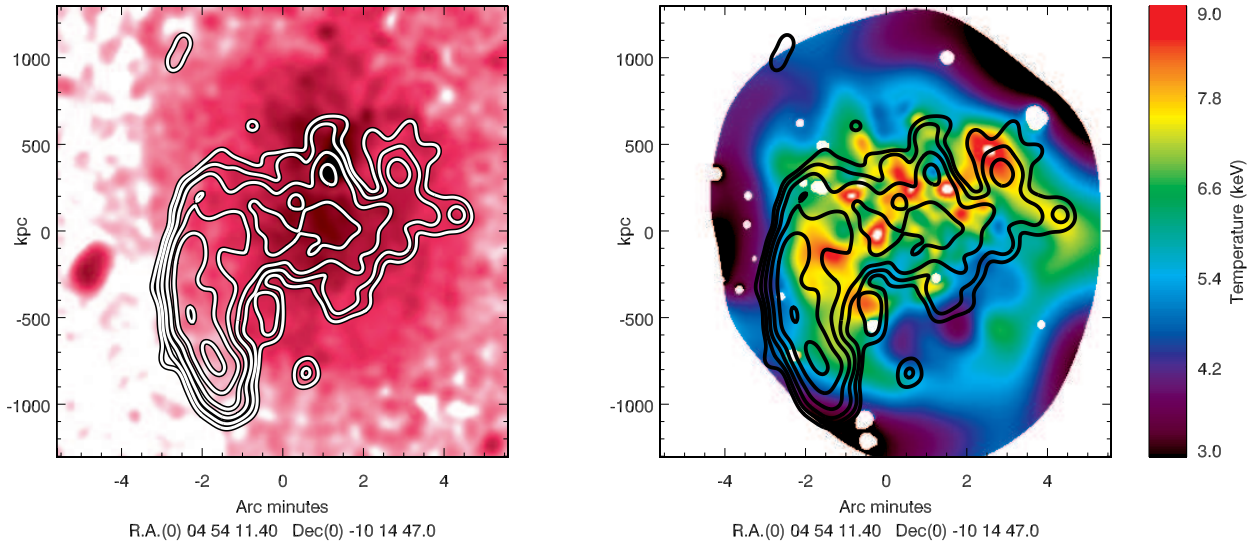}}
  \caption{{ Left panel:} Anisotropic details in the X-ray emissivity (same as \fig\ref{xmaps_fig}) overlaid with
  emissivity iso-contours in the 240 MHz radio band \citep[Giant Metrewave Radio Telescope;][]{Brunetti_08}. 
  { Right panel:} ICM temperature map overlaid with the same radio emissivity iso-contours as for the left figure.\label{x_radio_maps_fig}}
\end{figure*}

\begin{table*}[]
\caption{Density, temperature jumps and Mach numbers estimated across the shock fronts $\mathrm{S}_1$ and $\mathrm{S}_2$. \label{shock_params_tab}}
\begin{center}
\begin{tabular}{ccccc}
\tableline\tableline
&\multicolumn{2}{c}{Density estimates} & \multicolumn{2}{c}{Temperature estimates} \\
Shock front & Jump amplitude & Mach number & Jump amplitude & Mach number  \\
\tableline
$\mathrm{S}_1$ &$1.48^{+0.02}_{-0.11}$&$1.33^{+0.02}_{-0.08}$&$1.78^{+0.63}_{-0.38}$&$1.76^{+0.49}_{-0.35}$ \\
$\mathrm{S}_2$ &$2.64^{+0.13}_{-0.15}$&$2.42^{+0.19}_{-0.19}$&$4.47^{+8.06}_{-3.00}$&$3.40^{+3.69}_{-1.92}$ \\
\tableline
\tableline
\end{tabular}
\end{center}
\end{table*}

\subsection{Shock fronts propagation to the cluster outskirts\label{shock_fronts}}

The 2D gas brightness and temperature maps of \fig\ref{xmaps_fig} suggest the brightness jumps $\sone$ and $\stwo$ to be shock fronts propagating outwards the cluster center. Located at various distances from the cluster center, these two shock fronts might have been developed during two successive cluster collisions. 
In order to analyze ICM thermodynamics across these jumps, we
extracted the brightness and temperature profiles shown on
\fig\ref{a521_shocks_fig}, corresponding to the two sectors of
\fig\ref{shock_region_fig}. An estimation of the cluster photon counts
in each sector region is provided in \tab\ref{shock_region_tab}.
The brightness jumps $\sone$ and $\stwo$ exhibit the typical shape of a projected spherical density jump, convolved with the \xmm PSF.  We model the underlying gas density and temperature profiles as two step-like functions with common jump location, following equations (\ref{npne_cf_equ}) and (\ref{t3D_cf_equ}). A discussion about the validity of the assumption of the ICM spherical symmetry in the vicinity of the shock fronts is provided in the Appendix. The gas density and temperature distributions corresponding to this model are reported under the projected profiles on \fig\ref{a521_shocks_fig}. The 3D density and temperature jumps associated with these distributions are reported in \tab\ref{shock_params_tab}, with confidence intervals estimated from the 68 \% percentiles of a parameter sample matching several random realizations of the data set.

The direction of the temperature jumps is consistent
with the shock front interpretation. The cold front hypothesis would
instead imply a temperature increase across the jumps ($D_{\T} < 1$),
which is excluded by the data. Assuming two shocks propagating
outwards in the main cluster, one should be
able to estimate the shock Mach numbers from the Hugoniot-Rankine
density, temperature or pressure jump conditions across the
fronts. Such Mach number values are reported on
\tab\ref{shock_params_tab}. The Mach numbers independently estimated from the density
and temperature jumps are consistent with each other, though estimates
from the temperature jumps have larger uncertainties.  We
will hereafter use Mach number estimates for both shocks $\sone$ and $\stwo$,
from their density jumps: $\mathrm{M_\mathrm{\mathrm{S1}, \rho}} =
1.33^{+0.02}_{-0.08}$ and $\mathrm{M_\mathrm{\mathrm{S2}, \rho}} =
2.42\pm{0.19}$.

\section{Non-thermal ICM emission}{\label{sect:non_thermal_icm}}

A521 hosts a radio relic in its southeastern peripheral region, and a rare low-frequency giant radio halo. In order to investigate the interplay between thermal and non-thermal components of the ICM emission, the 240 MHz radio image obtained from observations performed at the Giant Metrewave Radio Telescope (GMRT) has been superimposed on the X-ray photon and ICM temperature maps of \fig\ref{x_radio_maps_fig}.

\subsection{The \a521 radio halo}

The \a521 radio halo has been discovered from low-frequency observations at the GMRT \citep[240, 325, 610 MHz,][see also \fig\ref{x_radio_maps_fig}]{Brunetti_08} and then studied in detail through a deep follow-up Very Large Array observation at 1.4 GHz \citep{Dallacasa_09}. Its very steep spectrum, with spectral index $\alpha \sim 1.9$ between 325 and 1400 MHz, suggests magneto-hydrodynamic turbulence to be responsible for the in-situ re-acceleration of the relativistic electrons \citep{Brunetti_08}. The radio halo is covering the cluster central region, exhibiting an E-W elongation and reaching the radio relic to the Southeast. When excluding the relic region, the halo appears spatially correlated with the cluster X-ray emission. There is an even better correlation between the radio brightness and the hottest regions of the ICM, --in particular, the radio brightness exhibits a quick drop across the $\sone$ shock.

The complex thermodynamics of the ICM in the cluster center hint at the possible origin of the turbulence that may re-accelerate non-thermal particles in the halo. The two cold fronts $\cfone$ and $\cftwo$ may have developed K-H instabilities at large angles from the main cluster collision axis. As suggested by the spatial correlation between shock-heated regions and the radio emission, turbulence may alternatively have been generated behind the two shocks $\sone$ and $\stwo$, now propagating to the cluster outskirts. In addition, the merger disturbance has likely generated turbulence within the two subcluster core remnants. 

\begin{figure}[t]
  %\resizebox{\hsize}{!}{\includegraphics{./figures/a521_xradio_dressler_610_ct7_cmyk.eps}}
 \resizebox{\hsize}{!}{\includegraphics{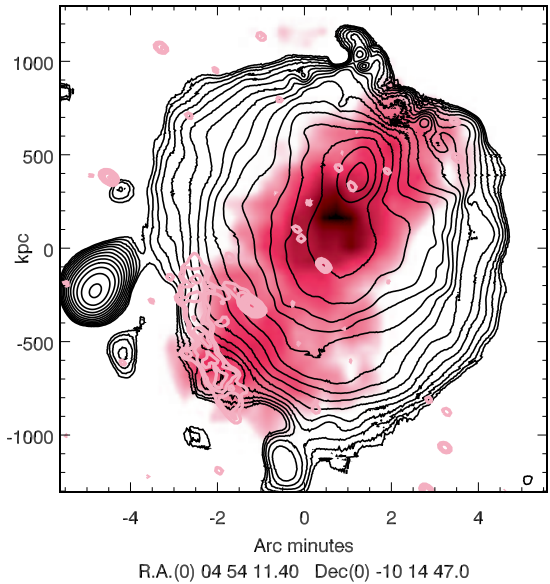}}

  \caption{Projected galaxy density distribution derived from photometric 
  	observations performed at the CFH telescope \citep[dark color indicate higher densities, Dressler algorithm, see][]{Ferrari_03}.  
	{ Pink iso-contours:} radio emissivity 
	in the 610 Mhz band \citep{Giacintucci_08}.  { Black iso-contours:} X-ray emissivity in the .5-2.5  $\kev$ band
	(Curvelet de-noising of the \xmm image). \label{galaxy_relic_fig}}
\end{figure}

\subsection{Shock front propagation and the radio relic}

\a521 has been known to host a SE radio relic observed at various
frequencies \citep[][hereafter, G08]{Ferrari_06,Giacintucci_06,Giacintucci_08}. 
As shown in G08, the integrated synchrotron radiation in the relic exhibit a power-law 
spectrum with spectral index $\alpha \simeq 1.5$ in the frequency range 235-4890 Mhz, 
with evidences of steepening of the radio spectrum with increasing distance from the eastern edge.
As further noted in G08, the outer edge of the radio relic coincides with the X-ray edge $\stwo$, 
which we have shown in this work to be a shock front propagating to the cluster outskirts.
As observed in several peripherical radio sources of galaxy clusters (see e.g. \citet{Bruggen_12} 
for a recent review), these facts support the shock electron (re)-acceleration to be at least partly responsible for the radio 
emission from the relic. Assuming diffuse shock acceleration for the origin of the emitting electrons, in the test particle approach the slope of the injection spectrum of cosmic rays is related to the shock Mach number, M, by \citep[][]{Blandford_87} $\delta_i = 2 (M^2+1)/(M^2-1)$. This leads to a spectrum of electrons in the downstream region with slope $\delta = \delta_i +1$ (implying a synchrotron spectral index
$\alpha = \delta_i/2$) taking into account radiative losses and
assuming stationary conditions. G08 thus predicted a shock propagation with Mach number, 
$\mathrm{M} \simeq 2.3$, from their measurement of $\alpha$. The steepening of the radio spectrum 
with increasing distance from the eastern edge further allowed them to predict a shock 
propagation to the cluster outskirts. The  propagation direction and Mach number of the shock front $\stwo$, 
$\mathrm{M_\mathrm{\rho}} = 2.4\pm0.2$ (cf. section \ref{shock_fronts}), are fully consistent with this hypothesis.

It is further worth noticing that the X-ray edge corresponding to the
shock front seems to extend in North--South direction more than the radio relic. 
A first interpretation for this limited extent of the relic might be that the shock would re-accelerate pre-existing relativistic electrons in the ICM. In this case the radio relic could reflect the spatial and energy distribution of the pre-existing electrons across the shock front. In line with this hypothesis, recent analyzes \citep{Kang07,Kang_11a,Kang_11b} suggest that the presence of pre-existing particles in addition to the thermal pool can significantly increase the average efficiency of the particle acceleration and the expected synchrotron emission at weak shocks (M $\le$ 3). Differences in extension between the shock and the radio relic might alternatively indicate some changes in the efficiency of electron acceleration changes along the shock front, possibly due to local variations of the Mach number \citep[e.g.,][]{Hoeft_08}. In this respect the radio relic in A521 is located at the extremity of the NW/SE major galaxy alignment evidenced in \citet[][see also
\fig\ref{galaxy_relic_fig}]{Ferrari_03}, where indeed recent accretion
of subcluster material may have produced inhomogeneities in the ICM.

\section{Discussion and conclusions}{\label{sect:discussion}}

A521 is a complex cluster system where optical analyzes have revealed 
at least three galaxy groups to the SE, and four groups to the NW including 
the cluster BCG group coinciding with the X-ray peak \citep{Ferrari_03}. The X-ray 
morphology of the BCG group suggests an infall along a NNW--SSE 
direction (projected onto the sky plane), which is slightly offset with respects to the major NW--SE 
galaxy alignment \citep{Ferrari_06}. The cluster atmosphere exhibits various 
brightness and temperature edges associated with cold fronts and shock fronts, that our \xmm data
revealed. 

The main two interacting gas components in the central region of this system are separated by a region of gas with lower density, higher temperature and entropy. We interpret this feature a flow of high-entropy gas being squeezed by two converging subcluster cores that are delimited by cold fronts. We suggest this high entropy gas to have been heated by shocks formerly developed when the two gas components started to interact. One of these shocks is currently observed to the South of the main component, with Mach number $\mathrm{M_\mathrm{\rho}} = 1.33^{+0.02}_{-0.08}$.  The hot gas region separating the two interacting components appears spatially correlated with the cluster radio halo. The development of turbulence in the hot gas flowing between the two cool cores may be responsible for high energy electron re-acceleration, yielding the radio halo emission. Merger shock propagation and/or cold fronts 
may have contributed to the development of these instabilities.

A shock front is observed at the Southeast cluster outskirt. An X-ray
brightness edge there has been hinted at by Chandra data (G08), though
the statistical significance was marginal. The orientation of this
shock front and its large distance from the cluster center suggest
that it is associated with a cluster collision that has occurred prior
to the current two-component interaction. Our Mach number for this shock, 
$\mathrm{M_\mathrm{\rho}} = 2.4\pm0.2$, is consistent with that
expected from the spectrum of the radio relic in G08 under the assumption 
of Fermi I acceleration mechanism. As observed in X-ray follow-ups of other 
radio relics --A3667, \citet{Finoguenov_10}; RXCJ1314.4-2515, \citet{Mazzotta_11}--,
its detection supports the shock electron (re)-acceleration to be at least partly responsible 
for the radio emission from the relic. The detection of a polarization of the relic 
would be an additional support for this process, complementary to the extension of its
synchrotron spectrum to very high radio frequencies, and to evidences for spectral 
steepening downstream to the shock (G08). Delimited by the shock front, the radio relic 
seems however to subtend only a fraction of the shock front. Differences in the spatial extent 
of a radio relic and its companion shock front have also been observed in the colliding cluster 
RXCJ1314.4--2515 \citep{Mazzotta_11}, where a radio relic seems to be confined to a small 
section of the shock front presumably distorted by a nonuniform gas flow. These
differences may thus reflect variations of the efficiency of particle
acceleration across the shock that could be driven by local variations
of the Mach number and shock velocity. Deeper X-ray or SZ
observations may enable us to investigate this hypothesis, though the
present XMM-Newton image does not seem to evidence any strong variation in the
amplitude of the surface brightness edge, and thus in the shock Mach
number. An alternative hypothesis is that the radio relic would reveal
us local inhomogeneities in the properties of the pre-existing
relativistic electrons, that would be re-accelerated by the shock
passage. The observed connection between the radio halo and the relic
may suggests that pre-existing relativistic electrons have
first been accelerated by turbulent gas motions responsible for the
radio halo emission, then re-accelerated at the shock front.

\acknowledgments
We thank the reviewer for her/his constructive comments and suggestions aiming at improving the manuscript.
H.B. thanks the Harvard-Smithsonian Centre for Astrophysics, where this work has been initiated, 
for its hospitality. We thank Chiara Ferrari for providing us a map of the projected galaxy density distribution 
in A521, derived from photometric observations performed at the CFH telescope. 
This work is based on observations obtained with XMM-Newton, an ESA science mission funded by 
ESA Member States and the USA (NASA). H.B and P.M acknowledge support by grants NASA grant 
NNX09AP45G and  NNX09AP36G grant ASI-INAF I/088/06/0 and ASI-INAF I/009/10/0. S.G. acknowledges 
the support of NASA through Einstein Postdoctoral Fellowship PF0-110071 awarded by the Chandra X-ray Center, 
which is operated by the Smithsonian Astrophysical Observatory. G.B. acknowledges partial support from
PRIN-INAF2009

\appendix

\begin{figure*}[t]
  %\resizebox{\hsize}{!}{\includegraphics{./figures/shock_asphericity.eps}}
  \resizebox{\hsize}{!}{\includegraphics{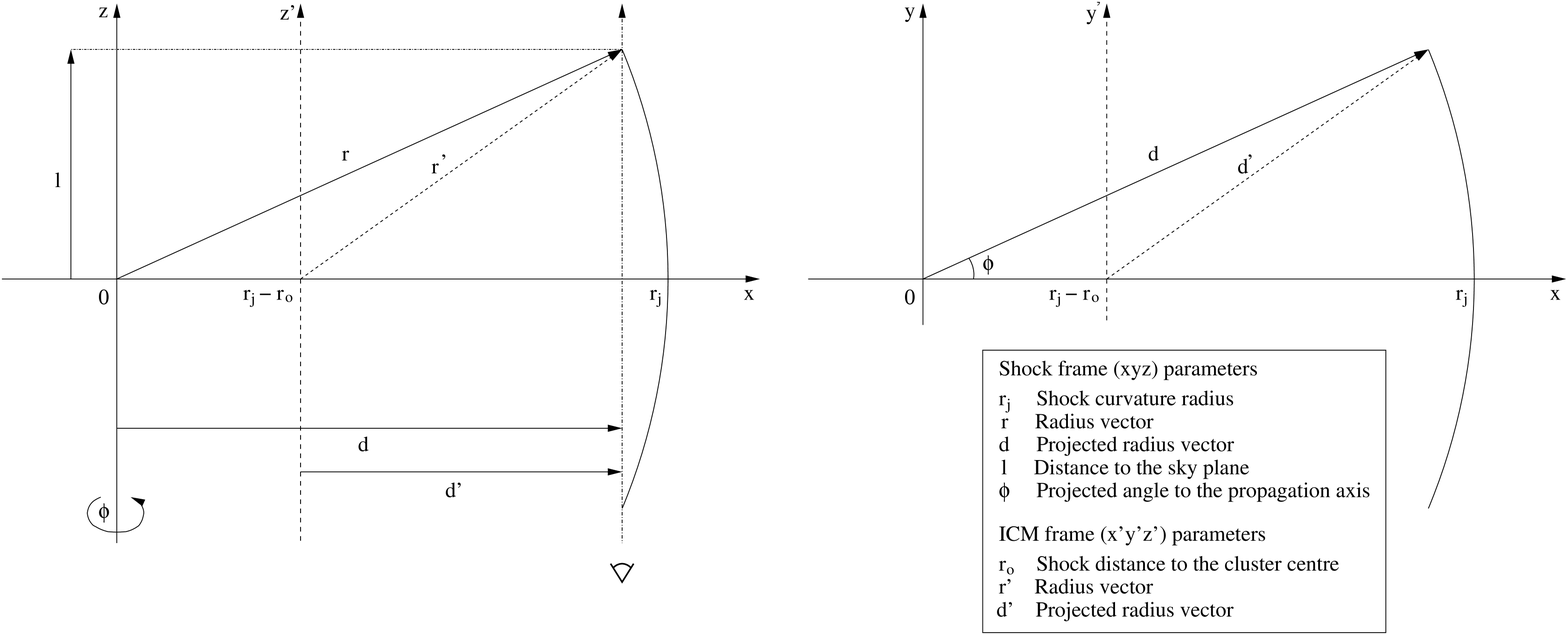}}
  \caption{ICM geometry across the shocks. \textit{Left:} Cluster volume cut along the line of sight.  \textit{Right:} Cluster volume cut within the sky plane. \label{shock_asphericity_fig}}
\end{figure*}

\begin{table*}[]
\caption{ICM asphericity parameters across the shock fronts S1 and S2. \label{icm_asphericity_tab}}
\begin{center}
\begin{tabular}{lcccccc}
\tableline\tableline
Shock & ICM asphericity        & ICM slope & Shock curvature  & Density jump     & Temperature jump & Mach number \\
      &$\left(1-r_o/r_j\right)$ & $(\eta)$  & ($\mathrm{r}_j$) & ($\mathrm{D}_n$) & ($\mathrm{D}_\mathrm{T}$) & (derived from $\mathrm{D}_n$) \\
\tableline
$\mathrm{S}_1$&$0.00^{+0.00}_{-0.00}$&$0.70^{+0.29}_{-0.50}$&$1.52^{+0.01}_{-0.01}$&$1.39^{+0.14}_{-0.07}$&$1.80^{+0.70}_{-0.38}$&$1.26^{+0.10}_{-0.05}$ \\
&$0.50^{+0.00}_{-0.00}$&$0.51^{+0.16}_{-0.34}$&$1.52^{+0.01}_{-0.01}$&$1.36^{+0.15}_{-0.07}$&$1.75^{+0.62}_{-0.36}$&$1.24^{+0.11}_{-0.05}$ \\
&$0.45^{+0.06}_{-0.18}$&$0.14^{+0.12}_{-0.02}$&$1.52^{+0.01}_{-0.01}$&$1.54^{+0.01}_{-0.06}$&$1.78^{+0.66}_{-0.36}$&$1.37^{+0.00}_{-0.05}$ \\
\tableline
$\mathrm{S}_2$&$0.00^{+0.00}_{-0.00}$&$0.00^{+0.26}_{-0.00}$&$2.25^{+0.02}_{-0.00}$&$2.71^{+0.14}_{-0.12}$&$4.50^{+4.50}_{-2.92}$&$2.51^{+0.22}_{-0.16}$ \\
&$0.50^{+0.00}_{-0.00}$&$0.00^{+0.10}_{-0.00}$&$2.25^{+0.02}_{-0.01}$&$2.73^{+0.25}_{-0.19}$&$4.36^{+4.36}_{-2.82}$&$2.54^{+0.41}_{-0.25}$ \\
&$0.51^{+0.02}_{-0.01}$&$0.00^{+0.13}_{-0.00}$&$2.26^{+0.01}_{-0.01}$&$2.74^{+0.11}_{-0.14}$&$3.95^{+4.45}_{-2.41}$&$2.55^{+0.17}_{-0.20}$ \\
\tableline
\end{tabular}
\end{center}
\end{table*}

%\section{Appendix material}

The ICM density and temperature distributions intercepting the shocks $\sone$ and $\stwo$ has been
modelled in \part\ref{radial_profiles_sect} as two step-like functions, assuming the
shock center of curvature and the ICM centroid coincide to coincide with each other. 

The X-ray image of \fig\ref{xmaps_fig} seems however to show that the shock fronts $\sone$ and
S2 are less curved than the closest cluster brightness isophotes. To
investigate the systematic uncertainties inherent to our spherical
symmetry approximation, we alternatively tried to model the shock front
and the ICM density as two spherical distributions with distinct
centers. Assuming these two centers to be located in the plane of the
sky, the ICM emission measure is now expressed per volume unit, as:

\begin{equation}
  [n_p n_e](r) = \left\{ \begin{matrix}
    D_{n}^2 n_o^2~\left(\frac{r'}{r_o}\right)^{-\eta}, ~r<r_{j} \\
    n_o^2~\left(\frac{r'}{r_o}\right)^{-\eta}, ~r>r_{j}
  \end{matrix} \right.,
  \label{npne_sf_equ}
\end{equation}

where $r$ and $r'$ refers to the norm of each radius vector in the shock
and ICM frames, respectively. Introducing $d$ and $d'$, the projection of
these radius vectors onto the sky plane, a surface brightness profile
intercepting the shock is obtained from integration of \equ(\ref{npne_sf_equ}) 
along the line of sight:

\begin{equation}
\Sigma(x) = 2  \times \int\limits_{{\phi_{min}}}^{\phi_{max}} \int\limits_{0}^{\infty}\int\limits_{x-\delta x}^{x+\delta x} \Lambda(T(d,l)) [n_p n_e](d,d',l,\phi) dd\, dl\, d\phi,
 \label{sigma_sf_equ}
\end{equation}

where $d'$ is related to $d$ as a function of $r_o$, the distance separating the shock from the
center of the ICM distribution and $\phi$, the angle separating the
projected radius vector to the shock propagation axis ($d'= \sqrt{(r_j-r_o)^2-2 (r_j-r_o) d \cos(\phi) +
d^2}$, see also Fig. \ref{shock_asphericity_fig}).
In addition to an ICM density slope, $\nu$, and the shock
curvature radius, density and temperature jumps, $r_j$, $\mathrm{D}_n$ and 
$\mathrm{D}_\mathrm{T}$, respectively, the ICM emission measure thus 
depends on an asphericity parameter: $1-r_o/r_j$. 

We tried to invert $[n_p n_e](r)$ and its parameters from a minimization of
the $\chi^2$ distance separating $\Sigma(x)$ (\equ\ref{sigma_sf_equ}) from the X-ray surface
brightness profiles extracted across each shock front (see
\fig\ref{a521_shocks_fig}). Some of the searched parameters being degenerated with one
another, we first performed this inversion by fixing the asphericity
parameter to 0 and 0.5, corresponding to shocks located at distances
of $r_j$ and $2 \times r_j$ from the cluster center, respectively. We subsequently
left all parameters free to vary and report the results of our
measurements in \tab\ref{icm_asphericity_tab}, the confidence interval on each parameter
being estimated from the 68 \% percentiles of a parameter sample matching several random 
realizations of the data set. As expected, the shock curvature radius, density and temperature jumps
obtained when fixing the asphericity to 0 are consistent with their
estimates derived from the spherical model of section
\ref{radial_profiles_sect}. A marginal difference in the amplitude of
the density jump is still noticeable, since \equ(\ref{npne_sf_equ}) yields 
$\mathrm{D}_n=1.39^{+0.14}_{-0.07}$ as for shock $\sone$, while \equ(\ref{npne_cf_equ})
yields $\mathrm{D}_n=1.33^{+0.02}_{-0.08}$. This difference is probably related to the lack of any
variation of the ICM density slope at the shock crossing, following \equ(\ref{npne_sf_equ}). 
Fixing the aspericity to 0.5 instead of 0 also marginally affect
the density jump, essentially due to the degeneracy between the ICM
asphericity and density slope. This degeneracy is noticeable in the
case of $\sone$, the shock front observed with the highest
statistics. Leaving the ICM asphericity free to vary yields estimates
of 0.45 and 0.75 in the case of $\sone$ and $\stwo$, respectively, consistent 
with the shock curvature radii observed on the X-ray image of
\fig\ref{xmaps_fig}. The shock density, temperature jump and Mach
numbers derived from these various assumptions are in any case
consistent with one another, and with their estimates obtained from
the sperical model of section \ref{radial_profiles_sect}. Given
the limited statistics available, it is difficult to break
the degeneracy between the ICM asphericity, ICM density slope and
shock curvature radius in the vicinity of the shocks. For simplicity
purposes, we consequently adopted the spherical model of section
\ref{radial_profiles_sect} in order to derive the amplitudes of the
density jumps and Mach numbers of the two shocks $\sone$ and $\stwo$.

\vspace{.25cm}
\bibliography{a521_arXiv}
%\clearpage

\end{document}